# Subtyping for F-Bounded Quantifiers and Equirecursive Types

Neal Glew

Intel Labs, Santa Clara, CA
aglew@acm.org

## 1 Introduction

*Equirecursive types* consider a recursive type to be equal to its unrolling and have no explicit term-level coercions to change a term's type from the former to the latter or vice versa. This equality makes deciding type equality and subtyping more difficult than the other approach—*isorecursive types*, in which the types are not equal, but isomorphic, witnessed by explicit term-level coercions. Previous work has built intuition, rules, and polynomial-time decision procedures for equirecursive types for first-order type systems. Some work has been done for type systems with parametric polymorphism, but that work is incomplete (see below). This chapter will give an intuitive theory of equirecursive types for second-order type systems, sound and complete rules, and a decision procedure for subtyping.

Another interesting feature of type systems turns out to be quite related to equirecursive types. Canning et al. [CCH+89] introduced the idea of *F-bounded polymorphism*. In this form of polymorphism a type bound can mention the type being bounded. For example, it can require a type that has a method that returns an object of the type being bounded. This form of bound is useful for binary methods and in typing object encodings [Gle00]. Treating a type variable as being a subtype of its bound when that bound can refer to it is like treating a recursive type as being equal to its unrolling, and similar issues arise to formalising such type systems. This chapter will also treat F-bounded parameteric polymorphism, and give it an intuitive formalisation, sound and complete set of rules, and decision procedure for subtyping.

Amadio and Cardelli [AC93] were the first to investigate the equirecursive approach. They proposed the *tree interpretation* of recursive types, which is based on the idea of repeatedly unrolling recursive types into possibly-infinite trees. Two types are equal if their corresponding trees are the same; similarly, subtyping can be defined on trees and lifted to types. Amadio and Cardelli made these ideas precise, defined a set of rules for type equality and subtyping that are sound and complete, and provided an exponential-time decision procedure for equality and subtyping. Kozen et al. [KPS95] reduced this exponential time to quadratic time, by defining a notion of tree automata that generate trees just as types do and a construction from two tree automata that decides equality and subtyping. Both Amadio and Cardelli and Kozen et al. worked with first-order type systems. Colazzo and Ghelli [CG99] investigated a second-order type system with equirecursion. They gave a coinductive set of rules for subtyping and a decision procedure, but did not relate their rules to trees nor show soundness and completeness to any intuitive model. In previous work [Gle02a,Gle02b], I gave a tree model for second-order systems, a set of rules for equality of types, showed soundness and completeness of those rules, defined automata for the trees, and defined a construction on automata that gives a decision procedure for equality.[1] However, I did not address subtyping. Gauthier and Pottier [GP04] devised an $O(n \log n)$ algorithm to decide equality of second-order types with equirecursive types by reducing second-order types to canonical first-order types in a particular way such that the canonical types are equal exactly when the original types are equal; their approach also handles entailment of type equations and type unification problems with the same complexity.

This chapter investigates a second-order type system with F-bounded quantifiers and equirecursive types. First, it defines a notion of trees that provide an intuitive model for such types and defines an intuitive notion of subtyping on these trees. Next, it presents the types themselves and defines how these map to trees. Then it presents a set of type equality and subtyping rules and shows soundness and completeness of these rules with respect to the tree interpretation. Finally, it defines a notion of tree automata, how these generate trees, and a construction that decides subtyping in polynomial time.

Complete definitions and proofs of all the results in this chapter appear in a companion technical report [Gle12].

## 2 Binding Trees

I consider a system with top, bottom, function, and F-bounded forall quantified types. To model such types, I use possibly-infinite trees over these constructs and de Bruijn indices to model type variables. These trees are formulated in the standard way:

$$\begin{aligned} \mathit{Tree} = \{t : \{\mathsf{L},\mathsf{R}\}^* \rightharpoonup \mathbb{N} \cup \{\top, \bot, \rightarrow, \forall\} \mid \\ \epsilon \in \mathrm{dom}(t) \wedge (p\ell \in \mathrm{dom}(t) \Leftrightarrow t(p) \in \{\rightarrow, \forall\})\} \end{aligned}$$

Forall quantifiers are F-bounded, that means that $\forall$ binds a variable in each of its subtrees—in the left subtree, the bound, to refer to the type being bounded and in the right tree to refer to the quantified type.

Trees form a complete 1-bounded ultrametic space in the usual way. Some sample trees appear in Figure 1, and I overload the notation and use $\mathrm{var}(n)$, $\top$, $\bot$, $t_\mathsf{L}{\rightarrow}t_\mathsf{R}$, and $\forall t_\mathsf{L}.t_\mathsf{R}$ to denote appropriate trees. The subtree of $t$ along path $p$ is $\mathit{subtree}(t,p)$, the number of variables bound along the path from $p_1$ to $p_2$ (a suffix) is $\mathit{bind}(t, p_1 \rightarrow p_2)$, and shifting the free variables of a tree $t$ up by $n$ is $\mathit{shift}(t,n)$; their definitions are straightforward.

[1] In those papers I claimed that the algorithm could be made quadratic time. I have recently realised that the reasoning was incorrect, and thus that my papers demonstrated only an exponential-time algorithm. The arguments I give latter can be used to make an $O(n^4)$ algorithm from the ones in my previous work.

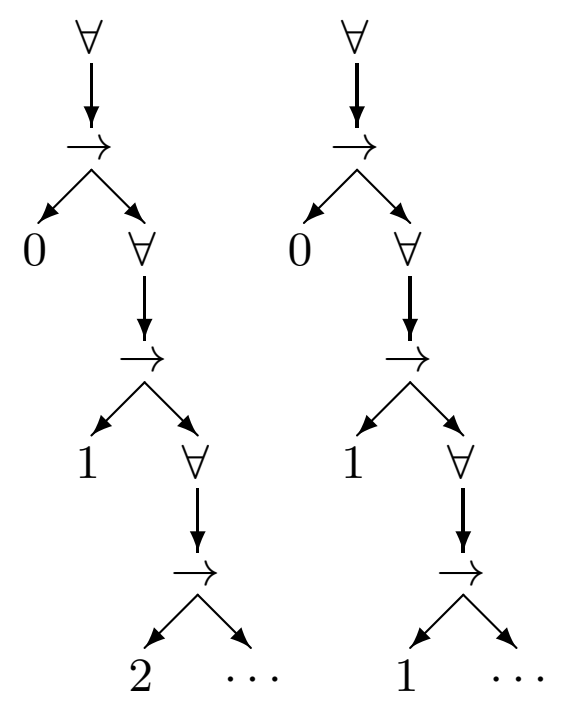

**Fig. 1.** Examples trees `t1`, on left, and `t2`, on right.

### 2.1 Regular Trees

Not all trees are generated by syntactic types. For first-order systems, regular trees—those with a finite number of subtrees—correspond exactly to the trees that can be generated. For the trees defined above, this is no longer the case. Consider tree `t1` in Figure 1 that is generated by $\forall\alpha.\mathsf{rec}\ \beta.\alpha \to \forall\beta'.\alpha$—all the de Bruijn indices intuitively represent the same thing, the variable bound by the $\forall$ at the top of the tree, but the actual numbers are all different. Now consider tree `t2` in Figure 1. It has only a finite number of subtrees, but in fact no type can generate it. It looks like the second and subsequent forall trees repeat each other, however the de Bruijn indices 1 in that tree refer to the quantifier in the previous iteration of the cycle rather than the current iteration, and types cannot generate such a structure. My previous work defined a rather complicated notion of regular binding trees. Since then I have discovered a better, more intuitive formulation.

The idea is that relevant subtrees are equal modulo appropriate changes to the de Bruijn indices. For example, in the tree for $\mathsf{rec}\ \alpha.\forall\beta \le \top.\alpha$, the tree itself is not equal to the right subtree, but the tree itself shifted by one is equal to the right subtree—one being the number of variables bound from the root to the right subtree. Generalising a bit, two subtrees at paths $p_1$ and $p_2$ with greatest common prefix $p$ that are intuitively equal cannot refer to any of the variables bound along the paths $p$ to $p_1$ or $p_2$ and will be equal modulo adjusting the de Bruijn indices according to the difference in the number of variables bound along these paths.

This idea can be formalised by defining an equivalence relation on the paths of a tree $t$, $\mathit{eqst}(t)$, that intuitively says which subtrees of $t$ are equal. First, $\mathit{forbid}(t, m)$ captures the notion that $t$ does not refer to the first $m$ free variables and shifts $t$ downwards by $m$. It is defined when there is no $p \in \mathrm{dom}(t)$ such that $t(p) = n$ and $\mathit{bind}(t, \epsilon \to p) \le n < \mathit{bind}(t, \epsilon \to p) + m$, otherwise it is undefined.

When it is defined:

$$forbid(t, m) = \lambda p. \begin{cases} n - m & t(p) = n \wedge n \geq bind(t, \epsilon \to p) \\ t(p) & \text{otherwise} \end{cases}$$

Now, $p_1$ $eqst(t)$ $p_2$ exactly when there exists $p$, $p'_1$, $p'_2$, and $q$ such that all of:

$$\begin{array}{l} p_1 = pp'_1 q \\ p_2 = pp'_2 q \\ forbid(subtree(t, pp'_1), bind(t, p \to pp'_1)) = \\ \quad forbid(subtree(t, pp'_2), bind(t, p \to pp'_2)) \end{array}$$

(The last one when both are defined and equal.) In other words, two paths represent equal subtrees of $t$ when they are common paths in two subtrees of $t$ that are equal when their de Bruijn indices are adjusted to the common prefix of the paths to those subtrees.

It is easy to prove that $eqst(t)$ is an equivalence relation on $\mathrm{dom}(t)$. A tree $t$ is a *regular binding tree* exactly when $[eqst(t)]$ is finite.

Each equivalence classes of $eqst(t)$ has a particular type constructor, as defined by the function:

$$NL(t, [p]_{eqst(t)}) = \begin{cases} \mathsf{fv}(n) & t(p) = n + bind(t, \epsilon \to p) \\ \mathsf{bv}([p']_{eqst(t)}, \ell) & p'\ell \leq p \wedge t(p'_1) = \forall \wedge t(p) = bind(t, p'\ell \to p) \\ \top & t(p) = \top \\ \bot & t(p) = \bot \\ \to & t(p) = \to \\ \forall & t(p) = \forall \end{cases}$$

It is easy to prove that this function is well defined and that $[p]_{eqst(t)}\ell = [p\ell]_{eqst(t)}$ when $NL(t, [p]_{eqst(t)}) \in \{\to, \forall\}$ is also well defined.

### 2.2 Subtyping

Intuitively, subtyping of trees should be as follows. Top should be a supertype of everything, bottom should be a subtype of everything, a variable should be a subtype of itself and its bound, a function tree is a subtype of another function tree when the argument trees are related contravariantly and the result trees are related covariantly, and similarly for forall trees. This could be made into a formal definition by using coinduction except for a couple of points. First, if the coinductive definition includes the condition that a variable is a subtype of any tree its bound is a subtype of then problems result. In particular, if free variable 0 is bounded by itself then under this definition free variable 0 is a subtype of any tree because of the coinduction—what we really want is induction for bounds. Second, comparing $\forall t_{1\mathsf{L}}.t_{1\mathsf{R}}$ to $\forall t_{2\mathsf{L}}.t_{2\mathsf{R}}$ requires selecting a bound for the variable bound by the $\forall$s to do the comparison of $t_{1\mathsf{L}}$ against $t_{2\mathsf{L}}$ and $t_{1\mathsf{R}}$ against $t_{2\mathsf{R}}$. The most general rule that is sound uses the tightest bound for the variable—this is $t_{2\mathsf{L}}$. However, that rule leads to an undecidable subtyping relation [Pie94]. The system considered here is actually a non-conversative extension,[2] but I believe that the undecidability still holds (but have not proven this yet). Therefore, to regain decidability, I will use the Kernel rule for forall—the bound is required to be invariant ($t_{1\mathsf{L}} = t_{2\mathsf{L}}$) and then the bound to use is the equal bound.

To formalise the above intuition, there are several pieces to build up, as mixing induction and coinduction is a little tricky. A bound set, ranged over by metavariable $\beta$, is a function from de Bruijn indices to trees, $\mathit{BSet} = \mathbb{N} \to \mathit{Tree}$. Shifting the free variables up and adding a bound $t$ for the new free variable 0 is $\mathit{shift}(\beta, t)$, and is straightforward to define. Tree promotion, a relation $\hookrightarrow$ on $\mathit{Tree} \times \mathit{BSet} \times \mathit{Tree}$, is defined as $\mathrm{var}(n) \hookrightarrow_\beta \beta(n)$ (no other trees are related by $\hookrightarrow_\beta$). The base subtyping proposition $\mathit{bst}(t_1, R, \beta, t_2)$ holds exactly when either:

- $t_1 = \mathrm{var}(n)$ and $t_2 = \mathrm{var}(n)$,
- $t_2 = \top$,
- $t_1 = \bot$,
- $t_1(\epsilon) = \to$, $t_2(\epsilon) = \to$, $\mathit{subtree}(t_2, \mathsf{L})\ R_\beta\ \mathit{subtree}(t_1, \mathsf{L})$, and $\mathit{subtree}(t_1, \mathsf{R})\ R_\beta\ \mathit{subtree}(t_2, \mathsf{R})$, or
- $t_1(\epsilon) = \forall$, $t_2(\epsilon) = \forall$, $\mathit{subtree}(t_1, \mathsf{L}) = \mathit{subtree}(t_2, \mathsf{L})$, and:

$$\mathit{subtree}(t_1, \mathsf{R})\ R_{\mathit{shift}(\beta, \mathit{subtree}(t_2, \mathsf{L}))}\ \mathit{subtree}(t_2, \mathsf{R})$$

A three place relation $R$ on $\mathit{Tree} \times \mathit{BSet} \times \mathit{Tree}$ is a partial subtyping exactly when $t_1\ R_\beta\ t_2$ implies that there exists $t_1'$ such that $t_1 \hookrightarrow_\beta^* t_1'$ and $\mathit{bst}(t_1', R, \beta, t_2)$. Subtyping for trees, $\leq$, is the union of all partial subtypings.

Subtyping satisfies several important properties justifying that the formal definitions do capture the right intuition.

**Theorem 1.** *Subtyping is a partial subtyping; $\leq_\beta$ is a preorder on Tree for any $\beta \in \mathit{BSet}$; $t_1 \leq_\beta t_2$ if and only if one of the following holds:*

- $t_1 = \mathrm{var}(n)$ *and* $t_2 = \mathrm{var}(n)$,
- $t_1 = \mathrm{var}(n)$ *and* $\beta(n) \leq_\beta t_2$,
- $t_2 = \top$,
- $t_1 = \bot$,
- $t_1 = t_{11} \to t_{12}$, $t_2 = t_{21} \to t_{22}$, $t_{21} \leq_\beta t_{11}$, *and* $t_{12} \leq_\beta t_{22}$, *or*
- $t_1 = \forall t_3.t_4$, $t_2 = \forall t_3.t_5$, *and* $t_4 \leq_{\mathit{shift}(\beta, t_3)} t_5$.

### 2.3 Characterising Subtyping

A (tree) subtyping problem is a triple $(t_\mathsf{L}, \beta, t_\mathsf{R})$ where $t_\mathsf{L} \leq_\beta t_\mathsf{R}$ might or might not hold. The definition of subtyping says that after promoting the subtype to its bound some finite number of times the two trees have to match in a certain sense. In particular, two trees match when they are the same de Bruijn index, the supertype is top, the subtype is bottom, both trees are functions, or both trees are forall quantifiers and their respective left subtrees are equal. With this definition, subtyping requires that after a finite number of promotions of the subtype the two trees must match and furthermore, if they match because they are functions then the respective left subtrees must be contravariantly related and the respective right subtrees must be covariantly related, and if they match because they are forall quantifiers then the respective right subtrees must be covariantly related. For these various subtrees we can repeat this process, finding matching trees for them, and so on. Thus for some set of paths we get trees that match if the original subtyping held.

[2] The system considered by Pierce does not include recursive types of either flavour, but the rules in this paper could be used to define a more permissive subtyping relation that is still sound for Pierce's system—in essence certain subtyping that has an infinite derivation in Pierce's rules would be allowed rather than rejected.

We can formalise this idea as follows. A subproblem of *stp* will be a triple $(t_{\mathsf{L}}, \beta, t_{\mathsf{R}})$ where $t_{\mathsf{L}}$ is the current subtype, $t_{\mathsf{R}}$ is the current supertype, and $\beta$ is the current bound set. For a subtyping problem *stp*, the initial subproblem for path $\epsilon$ is simply *stp*. If the initial subproblem for path $p$ is $(t_{\mathsf{L}}, \beta, t_{\mathsf{R}})$ and there is a $t'_{\mathsf{L}}$ such that $t_{\mathsf{L}} \hookrightarrow^*_\beta t'_{\mathsf{L}}$ and $t'_{\mathsf{L}}$ and $t_{\mathsf{R}}$ match then the final subproblem for path $p$ is $(t'_{\mathsf{L}}, \beta, t_{\mathsf{R}})$; otherwise the final subproblem for path $p$ fails. If the final subproblem for path $p$ is $(t_{\mathsf{L}}, \beta, t_{\mathsf{R}})$ and matches because both trees are functions then the initial subproblem for path $p\mathsf{L}$ is $(\mathit{subtree}(t_{\mathsf{R}}, \mathsf{L}), \beta, \mathit{subtree}(t_{\mathsf{L}}, \mathsf{L}))$ and the initial subproblem for path $p\mathsf{R}$ is $(\mathit{subtree}(t_{\mathsf{L}}, \mathsf{R}), \beta, \mathit{subtree}(t_{\mathsf{R}}, \mathsf{R}))$. Similarly, if both trees are forall quantifiers then the initial subproblem for path $p\mathsf{R}$ is $(\mathit{subtree}(t_{\mathsf{L}}, \mathsf{R}), \mathit{shift}(\beta, \mathit{subtree}(t_{\mathsf{L}}, \mathsf{L})), \mathit{subtree}(t_{\mathsf{R}}, \mathsf{R}))$. Thus any subtyping problem has a prefixed closed set of paths containing $\epsilon$ of initial and final subproblems.

The characterisation of subtyping is the following theorem.

**Theorem 2.** *$t_{\mathsf{L}} \leq_\beta t_{\mathsf{R}}$ if and only if all final subproblems of $(t_{\mathsf{L}}, \beta, t_{\mathsf{R}})$ do not fail.*

We can go further than just these definitions however. Each tree that appears in a subproblem comes, in some sense, from either of the original trees or one of the bounds. De Bruijn indices that are bounded by themselves (a trivial bound) are not interesting, so a tree identifier for a subtyping problem $(t_{\mathsf{L}}, \beta, t_{\mathsf{R}})$ is either $\mathsf{L}$, $\mathsf{R}$, or $n$ where $\beta(n) \neq \mathrm{var}(n)$. A node identifier is a tree identifier and a sequence of $\mathsf{L}$, $\mathsf{R}$, and $\mathsf{S}$s specifying to take the left subtree, right subtree, or shift by one starting from that tree. Any subproblem of *stp* can be represented as a triple $(\mathit{ni}_{\mathsf{L}}, \mathit{nis}, \mathit{ni}_{\mathsf{R}})$ where $\mathit{ni}_{\mathsf{L}}$ is a node identifier representing the subtype, $\mathit{ni}_{\mathsf{R}}$ is a node identifiers representing the supertype, and *nis* is a sequence of node identifiers representing the bounds of the binding variables that have been opened up. We can inductively define two partial maps that compute the node identifier representations of the initial and final subproblems or $\mathsf{F}$ for a final subproblem that fails.

Each node identifier maps to an equivalence class of the equivalence of subtrees of the tree that it comes from. If there are only finitely many de Bruijn indices with non-trivial bounds, each of those is a regular binding tree, and the original subtype and supertype trees are regular binding trees, then the node identifiers map to a finite set of equivalence classes. So the pair of the current subtype and supertype node identifiers map to a finite set. Thus for any sufficiently long path for which there are initial or final subproblems there will be a repeat of this pair. This property is the key to showing completeness of the subtyping rules, it will be used to cut off the proof of subtyping to make a finite derivation, as we shall see.

This characterisation of subtyping is also used to build an algorithm for deciding subtyping. Essentially, the algorithm constructs a deterministic finite-state automata that searches for paths for which the final subproblem fails. If it fails to find such a path, that is, its language is empty, then the subtyping holds. It uses just the equivalence classes of the node identifiers as well as some information about which binder equivalence classes correspond—but this information is finite too if there are finitely many equivalence classes. The algorithm is able to do promotion, matching, and determining subtrees with just this information. If there are a finite number of equivalence classes (as will be the case for regular binding trees and finite number of non-trivial bounds) then the search space is finite and the algorithm is a decision procedure.

## 3 Types

Now I will define the syntactic type system, map it to trees, and then give a sound and complete set of type equality and subtyping rules.

Let *Var* be a some set of type variables ranged over by metavariable $\alpha$. The set of types, *Type*, ranged over by metavariables $\tau$ and $\sigma$ is defined by this grammar:

$$\tau ::= \alpha \mid \top \mid \bot \mid \tau_1 \to \tau_2 \mid \forall \alpha \leq \tau_1 . \tau_2 \mid \mathsf{rec}\ \alpha . \tau$$

subject to the requirement that in $\mathsf{rec}\ \alpha.\tau$, $\tau$ is syntactically contractive in $\alpha$, written $\tau \downarrow \alpha$. The latter is defined by induction on the structure of $\tau$ as follows:

$$\begin{array}{ll}
\alpha' \downarrow \alpha & \Leftarrow \alpha \neq \alpha' \\
\top \downarrow \alpha & \\
\bot \downarrow \alpha & \\
\tau_1 \to \tau_2 \downarrow \alpha & \\
\forall \alpha' \leq \tau_1 . \tau_2 \downarrow \alpha & \\
\mathsf{rec}\ \alpha'.\tau \downarrow \alpha & \Leftarrow \alpha = \alpha' \vee \tau \downarrow \alpha
\end{array}$$

### 3.1 Mapping Types to Regular Binding Trees

Types map to trees given trees for the free variables. An environment, ranged over by metavariable $\eta$, maps type variables to trees, $\mathit{Env} = \mathit{Var} \to \mathit{Tree}$. An environment is distinguishing if it maps type variables injectively to $\{\mathrm{var}(n) \mid n \in \mathbb{N}\}$ (note this is weaker than my previous work, which required bijectivity). Shifting the free variables of an environment $\eta$ up and mapping $\alpha$ to the new free variable 0 is $\mathit{shift}(\eta, \alpha)$ and is straightforward to define.

The meaning of a type in an environment is a tree defined by induction on the type as follows:

$$\begin{aligned}
\mathit{treeof}(\alpha)_\eta &= \eta(\alpha)\\
\mathit{treeof}(\top)_\eta &= \top\\
\mathit{treeof}(\bot)_\eta &= \bot\\
\mathit{treeof}(\tau_1{\to}\tau_2)_\eta &= \mathit{treeof}(\tau_1)_\eta{\to}\mathit{treeof}(\tau_2)_\eta\\
\mathit{treeof}(\forall\alpha \le \tau_1.\tau_2)_\eta &= \forall \mathit{treeof}(\tau_1)_{\mathit{shift}(\eta,\alpha)}.\mathit{treeof}(\tau_2)_{\mathit{shift}(\eta,\alpha)}\\
\mathit{treeof}(\mathsf{rec}\ \alpha.\tau)_\eta &= \mathrm{fix}(\lambda t.\mathit{treeof}(\tau)_{\eta\{\alpha\mapsto t\}})
\end{aligned}$$

where $\mathrm{fix}(f)$ is the unique fixed point of a contractive function $f$ on trees (complete ultrametic spaces have unique fixed points for contractive functions)—the definition here is well defined as it is easy to prove that syntactic contractivity implies contractivity.

The meaning of a type is a regular binding tree and any regular binding tree is the meaning of some type.

**Theorem 3.** *If $\eta$ maps type variables to regular binding trees then $\mathit{treeof}(\tau)_\eta$ is a regular binding tree. If $t$ is a regular binding tree and $\eta$ can generate $t$'s free de Bruijn indices (there exists an $\alpha$ such that $\eta(\alpha) = \mathrm{var}(n)$ for each $n$ such that $t(p) = n + \mathit{bind}(t, \epsilon \to p)$) then there is a type $\tau$ such that $\mathit{treeof}(\tau)_\eta = t$.*

### 3.2 Equality and Subtyping Rules

To motivate the subtyping rules, consider some particular problems. First, if $\alpha$ is bounded by $\top{\to}\alpha$ then should $\alpha$ be a subtype of $\mathsf{rec}\ \alpha'.\top{\to}\alpha'$? Intuitively, $\alpha$ is some set $A$ of functions that take any value to a value in $A$; similarly $\mathsf{rec}\ \alpha'.\top{\to}\alpha'$ is the set $B$ of all functions that take any value to a value in $B$; it seems that $A$ should be a subset of $B$, so the subtyping should hold. Using de Bruijn index 0 for $\alpha$ then these types map to the trees $\mathrm{var}(0)$ and $t_2 = \{(\mathsf{R}^*, \to), (\mathsf{R}^*\mathsf{L}, \top)\}$ with bound set $\beta$ such that $\beta(0) = \top{\to}\mathrm{var}(0)$. Let $R = \{(\mathrm{var}(0), \beta, t_2), (\top, \beta, \top)\}$. Then $R$ is a partial subtyping and so $\mathrm{var}(0) \le_\beta t_2$. If the subtyping rules are to be complete then clearly they must be able to derive that $\alpha$ bounded by $\top{\to}\alpha$ is a subtype of $\mathsf{rec}\ \alpha'.\top{\to}\alpha'$.

Second, consider an example that does not even involve recursive types. If $\alpha$ is bounded by $(\alpha{\to}\top){\to}\bot$ then should $\alpha$ be a subtype of $\alpha{\to}\top$? These types map to trees $\mathrm{var}(0)$ and $\mathrm{var}(0){\to}\top$ with $\beta$ such that $\beta(0) = (\mathrm{var}(0){\to}\top){\to}\bot$. Let $R = \{(\mathrm{var}(0), \beta, \mathrm{var}(0){\to}\top), (\bot, \beta, \top)\}$. Then $R$ is a partial subtyping so $\mathrm{var}(0) \le_\beta \mathrm{var}(0){\to}\top$ and the subtyping for the types above should be derivable with the subtyping rules.

Using the standard structural subtyping rules with the equality rules from Amadio and Cardelli to try to prove these subtypings results in a cycle—after some steps what needs to be proved is what we are trying to prove. Here is the attempt for the first subtyping (where $B = \alpha \leq \top{\to}\alpha$ and $\tau_2 = \mathsf{rec}\ \alpha'.\top{\to}\alpha'$):

$$\dfrac{\dfrac{}{B \vdash \alpha \leq \top{\to}\alpha} \quad \dfrac{\dfrac{\dfrac{}{B \vdash \top \leq \top} \quad B \vdash \alpha \leq \tau_2}{B \vdash \top{\to}\alpha \leq \top{\to}\tau_2} \quad \dfrac{\dfrac{\dfrac{}{\vdash \tau_2 = \top{\to}\tau_2}}{\vdash \top{\to}\tau_2 = \tau_2}}{B \vdash \top{\to}\tau_2 \leq \tau_2}}{B \vdash \top{\to}\alpha \leq \tau_2}}{B \vdash \alpha \leq \tau_2}$$

Notice though that the steps make some progress, in that they use the structural subtyping rule for function types at least once, so coinductive proofs would prove this subtyping. A specialised rule for recursive types could prove this derivation, but in the other example, we really need something like coinduction (where $B = \alpha \leq (\alpha{\to}\top){\to}\bot$):

$$\dfrac{\dfrac{}{B \vdash \alpha \leq (\alpha{\to}\top){\to}\bot} \quad \dfrac{B \vdash \alpha \leq \alpha{\to}\top \quad \dfrac{}{B \vdash \bot \leq \top}}{B \vdash (\alpha{\to}\top){\to}\bot \leq \alpha{\to}\top}}{B \vdash \alpha \leq \alpha{\to}\top}$$

I will present normal inductive rules that in the rules that make progress, namely the structural subtyping rules for function and forall quantified types, allow the conclusion to be assumed in proving the subterms to have the appropriate subtyping relationship. This modification of the standard rules is enough to get sound and complete rules with respect to the tree interpretation of types.

Subtyping assumptions, ranged over by metavariable $A$, are sets of pairs of types, which I will write in the form $\tau_1 \leq \sigma_1, \ldots, \tau_n \leq \sigma_n$. Subtyping bounds, ranged over by metavariable $B$, have the form $\alpha_1 \leq \tau_1, \ldots, \alpha_n \leq \tau_n$ where the $\alpha_i$ are distinct. The meaning of subtyping bounds in a distinguishing environment is a bound set and is defined as:

$$\mathit{treeof}(\alpha_1 \leq \tau_1, \ldots, \alpha_n \leq \tau_n)_\eta = \lambda n. \begin{cases} \mathit{treeof}(\tau_i)_\eta & \eta(\alpha_i) = \mathrm{var}(n) \\ \mathrm{var}(n) & \text{otherwise} \end{cases}$$

The rules for type equality and subtyping appear in Figure 2 and define two judgements. $\vdash \tau_1 = \tau_2$ asserts that types $\tau_1$ and $\tau_2$ are equal and $A; B \vdash \tau_1 \leq \tau_2$ asserts that $\tau_1$ is a subtype of $\tau_2$ under assumptions $A$ and bounds $B$. The equality rules are those of Amadio and Cardelli. The interesting rules are EQROLL and EQUNQ. The former asserts that a recursive type is equal to its unrolling. The latter asserts that recursive types are unique—more specifically that two types that satisfy the same syntactically contractive equation are equal. It is key to proving completeness of the equality rules with respect to the tree interpretation of types. The subtyping rules are also fairly standard. There are the usual reflexivity, transitivity, variable bound, top, and bottom rules. Rule STASSUME allows an assumption to be used. Rule STFUN is the usual structural subtyping rule except that the conclusion can be assumed while proving the the argument types are contravariantly related and the result types are covariantly related. Rule STALL is the Kernel rule for F-bounded forall quantified types, again where

$\boxed{\vdash \tau_1 = \tau_2}$

$$\frac{\vdash \tau_2 = \tau_1}{\vdash \tau_1 = \tau_2}\ \text{EQSYM} \qquad \frac{\vdash \tau_1 = \tau_2 \quad \vdash \tau_2 = \tau_3}{\vdash \tau_1 = \tau_3}\ \text{EQTRANS}$$

$$\frac{}{\vdash \alpha = \alpha}\ \text{EQVAR} \qquad \frac{}{\vdash \top = \top}\ \text{EQTOP} \qquad \frac{}{\vdash \bot = \bot}\ \text{EQBOT}$$

$$\frac{\vdash \tau_1 = \tau_2 \quad \vdash \sigma_1 = \sigma_2}{\vdash \tau_1 \to \sigma_1 = \tau_2 \to \sigma_2}\ \text{EQFUN} \qquad \frac{\vdash \tau_1 = \tau_2 \quad \vdash \sigma_1 = \sigma_2}{\vdash \forall \alpha \leq \tau_1 . \sigma_1 = \forall \alpha \leq \tau_2 . \sigma_2}\ \text{EQALL}$$

$$\frac{\vdash \tau_1 = \tau_2}{\vdash \mathsf{rec}\ \alpha . \tau_1 = \mathsf{rec}\ \alpha . \tau_2}\ \text{EQREC} \qquad \frac{}{\vdash \mathsf{rec}\ \alpha . \tau = \tau\{\alpha \mapsto \mathsf{rec}\ \alpha . \tau\}}\ \text{EQROLL}$$

$$\frac{\vdash \tau_1 = \sigma\{\alpha \mapsto \tau_1\} \quad \vdash \tau_2 = \sigma\{\alpha \mapsto \tau_2\} \quad \sigma \downarrow \alpha}{\vdash \tau_1 = \tau_2}\ \text{EQUNQ}$$

$\boxed{A; B \vdash \tau_1 \leq \tau_2}$

$$\frac{\vdash \tau_1 = \tau_2}{A; B \vdash \tau_1 \leq \tau_2}\ \text{STREF} \qquad \frac{A; B \vdash \tau_1 \leq \tau_2 \quad A; B \vdash \tau_2 \leq \tau_3}{A; B \vdash \tau_1 \leq \tau_3}\ \text{STTRANS}$$

$$\frac{\tau_1 \leq \tau_2 \in A}{A; B \vdash \tau_1 \leq \tau_2}\ \text{STASSUME} \qquad \frac{\alpha \leq \tau \in B}{A; B \vdash \alpha \leq \tau}\ \text{STBOUND}$$

$$\frac{}{A; B \vdash \tau \leq \top}\ \text{STTOP} \qquad \frac{}{A; B \vdash \bot \leq \tau}\ \text{STBOT}$$

$$\frac{\begin{array}{c} A' = A, \tau_1 \to \sigma_1 \leq \tau_2 \to \sigma_2 \\ A'; B \vdash \tau_2 \leq \tau_1 \\ A'; B \vdash \sigma_1 \leq \sigma_2 \end{array}}{A; B \vdash \tau_1 \to \sigma_1 \leq \tau_2 \to \sigma_2}\ \text{STFUN} \qquad \frac{\begin{array}{c} A' = A, \forall \alpha \leq \tau_1 . \sigma_1 \leq \forall \alpha \leq \tau_2 . \sigma_2 \\ \vdash \tau_1 = \tau_2 \\ A'; B, \alpha \leq \tau_2 \vdash \sigma_1 \leq \sigma_2 \\ \alpha \notin \mathit{fv}(A) \cup \mathit{fv}(B) \end{array}}{A; B \vdash \forall \alpha \leq \tau_1 . \sigma_1 \leq \forall \alpha \leq \tau_2 . \sigma_2}\ \text{STALL}$$

**Fig. 2.** Typing Rules

the conclusion can be assumed when proving that the body types are covariantly related.

We are mainly interested in judgements of the form $\emptyset; B \vdash \tau_1 \leq \tau_2$, which I will write simply as $B \vdash \tau_1 \leq \tau_2$—the assumption sets are really just for internal use to prove such judgements.

I proved the soundness and completeness of the equality rules in previous work [Gle02a]. I repeat those proofs for the system in this chapter in the companion technical report [Gle12], and I will use them in proving the soundness and completeness of the subtyping rules.

### 3.3 Soundness

If two types are subtypes then the trees that they map to are subtypes.

**Theorem 4.** *If $B \vdash \tau_1 \leq \tau_2$ and $\eta$ is distinguishing then $\mathit{treeof}(\tau_1)_\eta \leq_{\mathit{treeof}(B)_\eta} \mathit{treeof}(\tau_2)_\eta$.*

**Proof:** The proof uses a generalisation of subtyping on trees that takes assumptions into account, and is then by induction over the derivation of the subtyping judgement using a lemma that says that subtyping with assumptions has properties similar to those of the rules. The soundness of the equality rules is also used for Rule STREF. □

### 3.4 Completeness

If the trees two types map to are subtypes then the rules can derive that they are subtypes.

**Theorem 5.** *If* $\eta$ *is distinguishing and* $\mathit{treeof}(\tau_1)_\eta \leq_{\mathit{treeof}(B)_\eta} \mathit{treeof}(\tau_2)_\eta$ *then* $B \vdash \tau_1 \leq \tau_2$.

The proof is in the companion technical report [Gle12]. As previously mentioned, the key to the proof is the characterisation of subtyping. It states that no final subproblem fails. Each final subproblem can be represented using node identifiers and these node identifiers come from a finite set of equivalence classes, thus on any sufficiently long path there will be a repeat of which equivalence class is the subtype and which equivalence class is the supertype. For any node identifier of the initial and final subproblems the proof builds a canonical type. Then the proof shows that Rule STBOUND can mimic promotion to a bound and that if two node identifiers match then one of the Rules EQVAR and STREF, STTOP, STBOT, STFUN, or STALL can prove the required subtyping. For STALL the proof uses the fact that the left subtrees are equal and the completeness of the equality rules to show that the bound types are equal; for the right subtree and both subtrees of STFUN the proof recurses to a longer path, for which there are initial and final subproblems. Finally at a repeat in the equivalence classes the proof shows that the requried subtyping is in the assumption set and uses Rule STASSUME. Finally, in various places the proof needs to show that the canonical types match up to other types, which it does by showing that they generate the same trees and by using the completeness of the equality rules; Rule STTRANS is used to combine everything together. The proof is just going through all the details of the above sketch.

## 4 Binding-Tree Automata

This section defines a notion of tree automata that generate trees and a construction that determines subtyping—it takes two tree automata to a DFA whose language is empty exactly when the subtyping relation holds.

A binding-tree automata is a quadruple $(Q, i, \delta, \mathit{lf})$ such that $Q$ is a finite set of states, $i \in Q$ is the initial state, $\delta : Q \times \{\mathsf{L}, \mathsf{R}\} \rightharpoonup Q$ is the transition function, $\mathit{lf} : Q \to \{\mathsf{fv}(n) \mid n \in \mathbb{N}\} \cup \{\mathsf{bv}(q, \ell) \mid q \in Q \wedge \ell \in \{\mathsf{L}, \mathsf{R}\}\} \cup \{\top, \bot, \to, \forall\}$ is the labelling function, $\delta(q, \ell)$ is defined if and only if $\mathit{lf}(q) \in \{\to, \forall\}$, and $\mathit{lf}(q) = \mathsf{bv}(q', \ell)$ only if $\mathit{lf}(q') = \forall$ and all paths from $i$ to $q$ go through $q'$ and on the last time through $q'$ they follow an $\ell$ edge. Intuitively, a tree automata takes as input a path through a tree and outputs the node label at the end of that path, which can be either a free variable (of the original tree), a bound variable (that bound by the last time through the identified state), top, bottom, function, or forall.

The tree that a tree automata generates is defined as follows:

$$
\begin{aligned}
\mathit{bind}(ql, \ell) &= \begin{cases} 1 & ql = \forall \\ 0 & ql \neq \forall \end{cases} \\
\mathit{shift}(f, q := n) &= \lambda q'. \begin{cases} 0 & q' = q \\ f(q') + n & q' \neq q \end{cases} \\
\hat{\delta}((q, n, f), \ell) &= (\delta(q, \ell), n + \mathit{bind}(\mathit{lf}(q), \ell), \mathit{shift}(f, q := \mathit{bind}(\mathit{lf}(q), \ell))) \\
\hat{\mathit{lf}}(q, n, f) &= \begin{cases} n + m & \mathit{lf}(q) = \mathsf{fv}(m) \\ f(q') & \mathit{lf}(q) = \mathsf{bv}(q', \ell) \\ \mathit{lf}(q) & \text{otherwise} \end{cases} \\
\mathit{treeof}(Q, i, \delta, \mathit{lf}) &= \lambda p. \hat{\mathit{lf}}(\hat{\delta}^*((i, 0, \lambda q.0), p))
\end{aligned}
$$

where $\delta^*$ is the obvious lifting of $\delta$ to sequences of edges. The formal definition just tracks enough information to determine the de Bruijn indices for the states labelled as free and bound variables, otherwise it follows the intuition above.

Binding-tree automata generate regular binding trees and all regular binding trees are generated by a binding-tree automata.

**Theorem 6.** *If $ta$ is a binding-tree automata then $\mathit{treeof}(ta)$ is a regular binding tree. If $t$ is a regular binding tree then there exists a binding-tree automata $ta$ such that $\mathit{treeof}(ta) = t$.*

### 4.1 Subtyping Algorithm

Now, I will define a construction that takes two binding-tree automata and produces a deterministic finite-state automata (in the usual sense), such that the DFA's language is empty if and only if the trees of the two binding-tree automata are in the subtyping relation. In particular, the DFA will search for paths that show that the two trees are not subtypes. The characterisation of subtyping tells us that such paths exists if and only if the trees are not subtypes. Most of the information for determining if states match after promotion is available from the labelling functions, but some additional information is needed. Specifically, the construction must track which binding states in one binding-tree automata correspond to which binding states in the other binding-tree automata, in order to determine if two bound variables match or not. This correspondence will be tracked by *partial bijections*, defined next.

A partial bijection $R$ between sets $A$ and $B$ is a set of pairs from $A$ and $B$ such that $(a_1, b_1) \in R$ and $(a_2, b_2) \in R$ implies that $a_1 = a_1$ if and only if $b_1 = b_2$. Partial bijection update is defined as: $R\{a \leftrightharpoons b\} = \{(a', b') \in R \mid a' \neq a \wedge b' \neq b\} \cup \{(a, b)\}$.

An automata bounds is a finite function from de Bruijn indices to binding-tree automata—de Bruijn indices without a bound are bounded by themselves.

An automata bounds generates a bound set as follows:

$$treeof(ba) = \lambda n. \begin{cases} treeof(ba(n)) & n \in \mathrm{dom}(ba) \\ \mathrm{var}(n) & n \notin \mathrm{dom}(ba) \end{cases}$$

The input to the construction, an (automata) subtyping problem, is a triple $(ta_{\mathsf{L}}, ba, ta_{\mathsf{R}})$ where $ta_{\mathsf{L}}$ and $ta_{\mathsf{R}}$ are binding-tree automata and $ba$ is an automata bounds. The construction will search over the various states of the various automata, so define a problem state of $(ta_{\mathsf{L}}, ba, ta_{\mathsf{R}})$ to be either $(\mathsf{L}, q)$ for $q$ a state of $ta_{\mathsf{L}}$, $(n, q)$ for $n \in \mathrm{dom}(ba)$ and $q$ a state of $ba(n)$, or $(\mathsf{R}, q)$ for $q$ a state of $ta_{\mathsf{R}}$. Define the transition function, $\delta$, and the labelling function, $lf$, for $(ta_{\mathsf{L}}, ba, ta_{\mathsf{R}})$ by lifting the underlying transition functions and labelling functions in the obvious way. The states of the DFA are quadruples $(q_1, \phi, q_2, R)$ where $q_1$ and $q_2$ are problem states, $\phi \in \{+, \circ, -\}$ is a variance ($+$ means that $q_1$ should be a subtype of $q_2$; $\circ$ means that $q_1$ should be equal to $q_2$; $-$ means that $q_1$ should be a supertype of $q_2$), and $R$ is a partial bijection between problem states.

I build the formal definition of the construction up in several pieces. First, I define how problem states are promoted, that is, if they are variables they are replaced with their bounds, as follows: $q \hookrightarrow_{(ta_{\mathsf{L}}, ba, ta_{\mathsf{R}})} (n, i)$ if $lf(q) = \mathsf{fv}(n)$, $n \in \mathrm{dom}(ba)$, and $i$ is the initial state of $ba(n)$; and $q \hookrightarrow_{stp} \delta(q', \mathsf{L})$ if $lf(q) = \mathsf{bv}(q', \ell)$.

Second, a DFA state matches, $matches_{stp}(q_1, \phi, q_2, R)$, exactly when one of the following holds:

- $lf(q_1) = lf(q_2) = \mathsf{fv}(n)$,
- $lf(q_1) = \mathsf{bv}(q_1', \ell)$, $lf(q_2) = \mathsf{bv}(q_2', \ell)$, and $(q_1', q_2') \in R$,
- $lf(q_1) = \top$ and $\phi = -$, $lf(q_2) = \top$ and $\phi = +$, or $lf(q_1) = lf(q_2) = \top$,
- $lf(q_1) = \bot$ and $\phi = +$, $lf(q_2) = \bot$ and $\phi = -$, or $lf(q_1) = lf(q_2) = \bot$,
- $lf(q_1) = lf(q_2) = \rightarrow$, or
- $lf(q_1) = lf(q_2) = \forall$.

Intuitively, a state matches if the base subtyping proposition holds for the nodes represented by that state. Using it and the notion of promotion, I can define a function that promotes a DFA state if possible to a matching DFA state. In particular, define $promote_{stp}(q_1, +, q_2, R) = (q_1', +, q_2, R)$ where $q_1'$ is the first $q_1'$ such that $q_1 \hookrightarrow^*_{stp} q_1'$ and $matches_{stp}(q_1', +, q_2, R)$ or $q_1' = q_1$ if no such $q_1'$ exists; similarly, define $promote_{stp}(q_1, -, q_2, R) = (q_1, -, q_2', R)$ where $q_2'$ is the first $q_2'$ such that $q_2 \hookrightarrow^*_{stp} q_2'$ and $matches_{stp}(q_1, -, q_2', R)$ or $q_2' = q_2$ if no such $q_2'$ exists; define $promote_{stp}(q_1, \circ, q_2, R) = (q_1, \circ, q_2, R)$.

Third, the subtree of a DFA state along an edge is defined as follows:

$$\begin{aligned} subtree_{stp}((q_1, \phi, q_2, R), \mathsf{L}) &= (\delta(q_1, \mathsf{L}), -\phi, \delta(q_2, \mathsf{L}), R\{q_1 \leftrightharpoons q_2\}) & lf(q_1) = \rightarrow \\ subtree_{stp}((q_1, \phi, q_2, R), \mathsf{L}) &= (\delta(q_1, \mathsf{L}), \circ, \delta(q_2, \mathsf{L}), R\{q_1 \leftrightharpoons q_2\}) & lf(q_1) = \forall \\ subtree_{stp}((q_1, \phi, q_2, R), \mathsf{R}) &= (\delta(q_1, \mathsf{R}), \phi, \delta(q_2, \mathsf{R}), R\{q_1 \leftrightharpoons q_2\}) & lf(q_1) \in \{\rightarrow, \forall\} \end{aligned}$$

(Where $-+ = -$, $-\circ = \circ$, and $-- = +$.) Intuitively it computes the state that corresponds to the left or right subproblem of a DFA state, updating the variance and binding correspondence in the appropriate way.

Finally, the subtype automata is defined as follows:

$$
\begin{aligned}
&\mathit{subtype}(ta_{\mathsf{L}}, ba, ta_{\mathsf{R}}) = \\
&\quad (Q \times \mathit{Var} \times Q \times (Q \leftrightharpoons Q), \\
&\quad\ \mathit{promote}_{stp}(q_{\mathsf{L}}, +, q_{\mathsf{R}}, \emptyset), \\
&\quad\ \lambda(q, \ell).\mathit{promote}_{stp}(\mathit{subtree}_{stp}(q, \ell)), \\
&\quad\ \{q \mid \neg \mathit{matches}_{stp}(q)\})
\end{aligned}
$$

where $Q$ is the set of problem states of $(ta_{\mathsf{L}}, ba, ta_{\mathsf{R}})$, $q_{\mathsf{L}}$ is the initial problem state of $ta_{\mathsf{L}}$, and $q_{\mathsf{R}}$ is the initial problem state of $ta_{\mathsf{R}}$.

The DFA so constructed has an empty language exactly when the subtyping relation holds.

**Theorem 7.** *$L(\mathit{subtype}(ta_{\mathsf{L}}, ba, ta_{\mathsf{R}})) = \emptyset$ if and only if $\mathit{treeof}(ta_{\mathsf{L}}) \leq_{\mathit{treeof}(ba)} \mathit{treeof}(ta_{\mathsf{R}})$.*

**Proof:** (Sketch) First the proof shows that for all the paths for which there are initial subproblems that the states computed by the subtype DFA (in some sense) generate the corresponding trees of the subproblem. If the subtyping holds then all the final subproblems do not fail. The proof then shows that those paths match in the subtype DFA and are thus not in the language. The only other paths the DFA considers are following a left edge from a forall matching subproblem, where the DFA switches to invariance and becomes an equality checker. Since to forall match the left subtrees must be equal, none of those paths will be in the language. Conversely if the language is empty then the proof shows that the left edge from forall matching states implies the left subtrees are equal and thus the corresponding final subproblem does not fail, and that all other final subproblems do not fail from their paths not being in the language. Thus by characterisation of subtyping the subtyping holds. The rest of the proof is just working through all the tedious details. □

Since determining if the language of a DFA is linear time, the construction provides an exponential time algorithm for deciding subtyping (at least of automata).

### 4.2 Polynomial Time Algorithm

The key to getting a polynomial time algorithm is that the binder correspondence information is only used in very limited ways in the subtype DFA, and so it can almost be ignored. For this section, fix a subtyping problem *stp*. Let $Q$ be the problem states of *stp*, *lf* the labelling function for problem states, and $\mathit{subtype}(stp) = (Q', i, \delta', F)$.

A triple is binder correspondence independent, $\mathit{bci}_{stp}(q_1, \phi, q_2)$, exactly when $\phi = +$ and $\mathit{lf}(q_2)$ is not a bound variable, $\phi = \circ$ and either $\mathit{lf}(q_1)$ or $\mathit{lf}(q_2)$ is not a bound variable, or $\phi = -$ and $\mathit{lf}(q_1)$ is not a bound variable; $\mathit{bci}_{stp}(q_1, \phi, q_2, R)$ exactly when $\mathit{bci}_{stp}(q_1, \phi, q_2)$. If $\mathit{bci}_{stp}(q_1, \phi, q_2)$ then $\mathit{matches}_{stp}(q_1, \phi, q_2, R_1)$ if and only if $\mathit{matches}_{stp}(q_1, \phi, q_2, R_2)$ for any $R_1$ and $R_2$. Then observe that if ${\delta'}^*(i, p\ell)$ is defined then ${\delta'}^*(i, p)$ is binder correspondence independent. Thus if $\delta'^*(i,p)$ is binder correspondence independent then determining if $p \in F$ can be done without tracking the binder correspondence at all.

Now consider $p$ such that $\delta'^*(i,p)$ is binder correspondence dependent. By definition $\delta'^*(i,p) = \mathit{promote}_{stp}(q_{\mathsf{L}}, \phi, q_{\mathsf{R}}, R)$ for some $q_{\mathsf{L}}$, $\phi$, $q_{\mathsf{R}}$, and $R$. The first three can be determined without tracking the binder correspondence. Consider what information is needed to determine if $p \in F$. Case 1, $\phi = \circ$: In this case $\mathit{lf}(q_{\mathsf{L}}) = \mathsf{bv}(q'_{\mathsf{L}}, \ell_{\mathsf{L}})$ and $\mathit{lf}(q_{\mathsf{R}}) = \mathsf{bv}(q'_{\mathsf{R}}, \ell_{\mathsf{R}})$ and $p \in F$ if and only $(q'_{\mathsf{L}}, q'_{\mathsf{R}}) \in R$. Case 2, $\phi = +$: In this case $\mathit{lf}(q_{\mathsf{R}}) = \mathsf{bv}(q'_{\mathsf{R}}, \ell)$. If there is no $q'_{\mathsf{L}}$ such that $(q'_{\mathsf{L}}, q'_{\mathsf{R}}) \in R$ then $p \in F$. If there is then $p \in F$ if and only if $q_{\mathsf{L}} \hookrightarrow^*_{stp} q'_{\mathsf{L}}$. Case 3, $\phi = -$: similar to the previous case.

My strategy for determining the above conditions is to compute facts of the form $q_1 \leftrightharpoons q_2$, $q \uparrow$, and $\uparrow q$ at the triples that are binder correspondence dependent. The meaning of $q_1 \leftrightharpoons q_2$ is that there is an $R$ possible at the triple with $(q_1, q_2) \in R$; similarly $q \uparrow$ means there is an $R$ possible at the triple with no $q'$ such that $(q, q') \in R$; and $\uparrow q$ means there is an $R$ possible at the triple with no $q'$ such that $(q', q) \in R$. Computing such facts is a simple dataflow problem. If $(q_1, \phi, q_2)$ is such that $\mathit{lf}(q_1) = \mathit{lf}(q_2) = \forall$ then $q_1 \leftrightharpoons q_2$ is generated, $q'_1 \leftrightharpoons q'_2$ is propagated if $q_i \neq q'_i$, $q \uparrow$ is propagated if $q \neq q_1$, $\uparrow q$ is propagated if $q \neq q_2$, $q_1 \leftrightharpoons q$ where $q \neq q_2$ is changed to $\uparrow q$, and $q \leftrightharpoons q_2$ where $q \neq q_1$ is changed to $q \uparrow$. Function states propagate all facts.

In summary, the polynomial time algorithm computes the triples that are possible and checks that the binder correspondence independent ones match and sets aside the binder correspondence dependent ones. Then it sets up and solves the dataflow problem outlined above. Finally it uses the computed facts at the binder correspondence dependent triples to determine if they match or not. The algorithm is at worst $O(n^4)$ as quadratic dataflow facts need to be propagated to quadratic nodes, and the other phases are at least as good. It might be possible to do better by exploiting the scoping requirements of binders, but I have not explored this possibility.

## 5 Discussion

To put all the pieces together, all we need to do is define a way to go from types to binding-tree automata. In my previous work I defined exactly such a transformation—in particular given a type $\tau$ and a distinguishing environment $\eta$ there is a binding-tree automata $\mathit{automataof}_{\eta}(\tau)$ constructable in linear time such that $\mathit{treeof}(\tau)_{\eta} = \mathit{treeof}(\mathit{automataof}_{\eta}(\tau))$. Combining that algorithm with the one in Section 4, gives a polynomial-time algorithm for deciding subtyping on the types themselves. The soundness and completeness of the rules and correctness of the algorithm means both that this algorithm is deciding subtyping according to the type rules, and that the type rules correspond to the tree interpretation of subtyping, which hopefully corresponds to our intuitive notion of what subtyping should be for the system under consideration.

**Theorem 8.** *If $\eta$ is distinguishing then:*

$$B \vdash \tau_1 \leq \tau_2$$
$$\Leftrightarrow$$
$$L(\mathit{subtype}(\mathit{automataof}_\eta(\tau_1), \mathit{automataof}_\eta(B), \mathit{automataof}_\eta(\tau_2))) = \emptyset$$

This chapter considered the Kernel rule for subtyping forall quantified types. As previously mentioned, that rule is not the most general rule that is sound for such types. I believe that the definition of subtyping for binding trees and the typing rules can be modified for the most general rule and the soundness and completeness theorems can still be proven, but I have not done this. Nothing in the proofs is critically dependent on the bound being invariant. The construction of a DFA for subtyping, however, is critically dependent on the bound being invariant. In particular, to augment the construction for the full rule first requires tracking which side is the tightest bound (easy to do), but also requires figuring out the binder correspondence after promoting to a bound, which requires saving the correspondence at the binding point leading to a linked list like correspondence information—no longer a finite set. Thus I believe this system is undecidable for similar reasons to full $F_{\leq}$.

There is another rule for subtyping forall quantified types that allows the bound to be contravariant but considers the variable unbounded in the body type. It is usually ignored as it leads to a lack of principal types. I believe that this rule could also be worked into my definition of trees, type rules, and automata construction—the variables bound by such quantifiers have no bounds, and so they act very similar to function types. I use a self quantifier in my object encoding that requires the body type to be covariant. As the body of the self quantifier is also the bound of the quantified variable, the typing rules cannot have an invariant bound. This variant with unbounded variables for checking the body is the most appropriate. The full rule is likely undecidable, and lack of principal types is avoided because the introduction form for self quantifiers includes a full type annotation, unlike for type lambdas that include only a partial type annotation.

Thus I believe that the results of this paper can be extended to handle other base types, first-order constructs of various flavours, other bound rules, and other second-order constructs like existentials and self quantifiers. Extending to higher-kinded systems is definitely future work and might not be possible. The first problem is that $F_\omega$ with equirecursive types has the simply-typed lambda calculus with general recursion at the type level, hence guaranteeing termination is probably a necessary first requirement.